\documentclass{jpsj-suppl}
\usepackage{txfonts} 
\usepackage{xspace}
\title{Radio detection of air showers with LOFAR and AERA}
\author{J\"org R. \textsc{H\"orandel}$^{1,2}$ 
on behalf of the LOFAR key science project Cosmic Rays and the
Pierre Auger Collaboration}

\inst{
$^{1}$Radboud University Nijmegen, Department of Astrophysics, IMAPP,
P.O. Box 9010, 6500 GL Nijmegen, The Netherlands, http://particle.astro.ru.nl\\
$^{2}$ Nikhef}

\email{j.horandel@astro.ru.nl}

\abst{
}

\kword{UHECR 2014, cosmic rays, air showers, radio emission, radio detection}

\def\lleft{{\sl left\xspace}}
\def\rright{{\sl right\xspace}}
\def\LLeft{{\sl Left\xspace}}
\def\RRight{{\sl Right\xspace}}
\def\deg{$^\circ$\xspace}
\def\Xmax{$X_{max}$}
\def\gcm2{g/cm$^2$}
\def\Cerenkov{\v{C}erenkov\xspace}

\def\cref#1{Chapt.\,\ref{#1}}
\def\Cref#1{Chapter~\ref{#1}}

\def\fref#1{Fig.\,\ref{#1}}

\begin{document}
\maketitle

\section{Introduction}
To understand the origin of high-energy cosmic rays is one of the open key
questions in astroparticle physics \cite{behreview,naganowatson}.  An inspiring
article, published in 2003 \cite{Falcke:2002tp} heralded the renaissance of
radio detection of extensive air showers with the ultimate goal to measure the
properties of cosmic rays with this technique and the pioneer experiments LOPES
\cite{radionature} and CODALEMA \cite{codalema} were initiated.  The
big success of these pathfinders stimulated further investigations of the radio
emission of air showers on larger scales, with installations such as Tunka-Rex
\cite{Kostunin:2013iaa}, AERA (Auger Engineering Radio Array) at the Pierre
Auger Observatory, and the LOFAR radio telescope.  Significant progress has
been achieved in the last decade \cite{Huege:2013eaa} and we now understand the
emission processes of the radio waves in the atmosphere.  Most of the emission
is due to the interaction of the shower with the magnetic field of the Earth,
which leads to a transverse current in the shower.  In addition to this
emission, the overabundance of electrons in the shower that are collected from
atmospheric molecules leads to a current in the direction of the shower.

\section{Radio Air-Shower Detectors}
\begin{figure}[t]\centering
\includegraphics[width=0.8\linewidth]{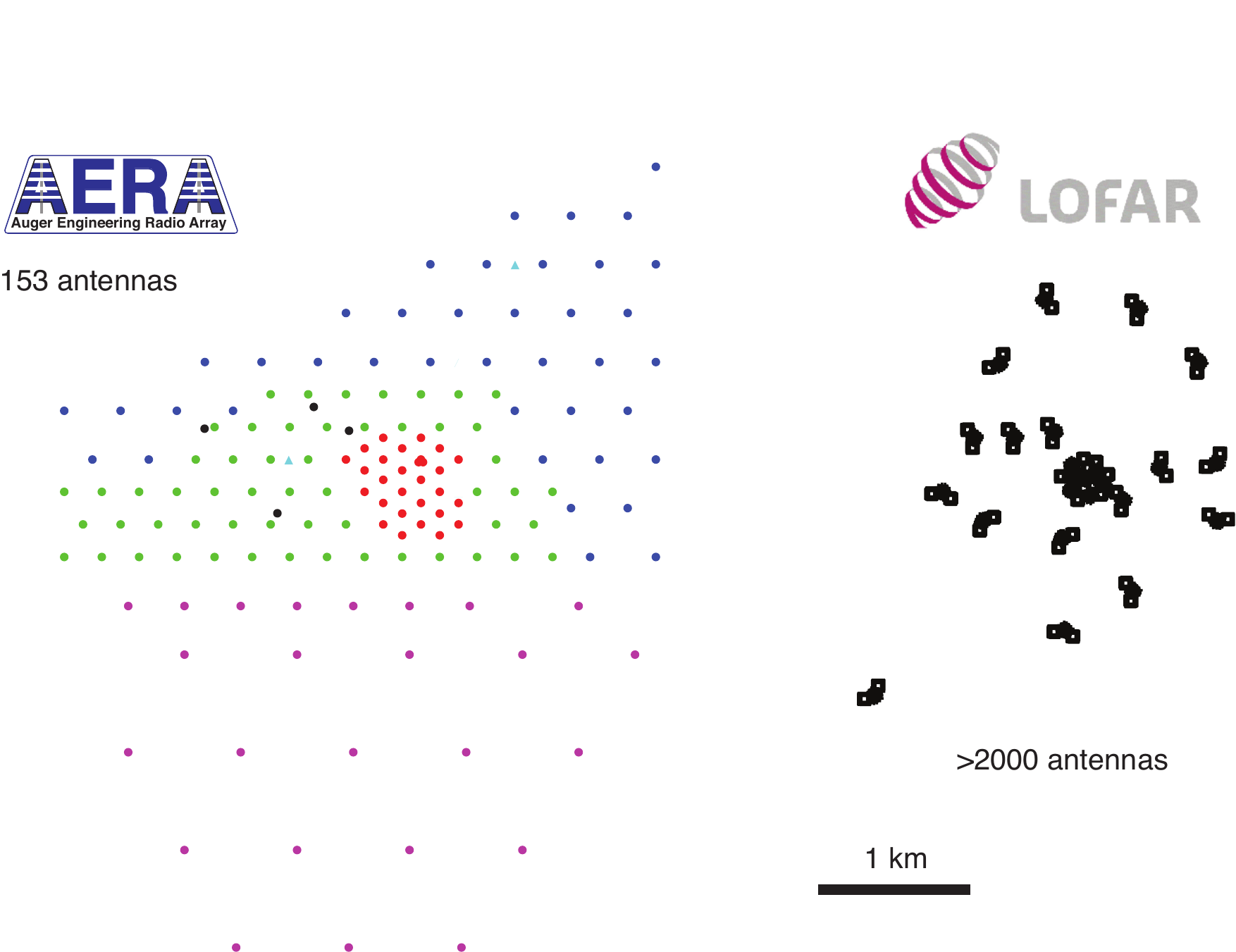}
\caption{Layout of AERA at the Pierre Auger Observatory and the dense core of
LOFAR -- drawn to scale.}
\label{layout}
\end{figure}

Two modern radio detectors for extensive air showers are the LOFAR radio
telescope (in particular, its dense core in the Netherlands) and AERA at the
Pierre Auger Observatory in Argentina.  The layout of the two set-ups is drawn
to scale in \fref{layout}.

\paragraph{LOFAR radio telescope}
The LOFAR key science project Cosmic Rays represents one of the six scientific
key objectives of LOFAR \cite{vanHaarlem:2013dsa, Schellart:2013bba}.
LOFAR is a digital radio telescope. Its antennas are spread over several
European countries and are used together for interferometric radio observations
in the frequency range of $10-240$~MHz. The density of antennas increases
towards the center of LOFAR, which is located in the Netherlands. Here, about
2400 antennas are clustered on an area of roughly 10~km$^2$. This high density
of antennas makes LOFAR the perfect tool to study features of the radio
emission created by extensive air showers.
Air-shower measurements are conducted based on a trigger received from an array
of scintillators (LORA) \cite{Thoudam:2014cfa,Thoudam:2015lba}, which results
in a read-out of the ring buffers that store the raw voltage traces per antenna
for up to 5~s.  LOFAR comprises two types of antennas, recording radio emission
in low-frequency band from 10 to 90 MHz and also in the high-frequency band
($110-240$~MHz).

\paragraph{AERA at the Pierre Auger Observatory}
AERA is a radio extension of the Pierre Auger Observatory. The Observatory is a
3000-km$^2$ hybrid cosmic-ray air-shower detector in Argentina
\cite{augerexp,Abraham:2009pm}, with an array of 1660 water-\Cerenkov detectors
and 27 fluorescence telescopes at four locations on the periphery
\cite{augernim}. The area
near the Coihueco fluorescence detector contains a number of low-energy
enhancements, including AERA.  AERA is located in a region with a higher
density of water-\Cerenkov detectors (on a 750 m grid) and within the field of
view of HEAT \cite{Meurer:2011ms}, allowing for the calibration of the radio
signal using “super-hybrid” air-shower measurements, i.e., recording
simultaneously the fluorescence light, the particles at the ground, and the
radio emission from extensive air showers.

Since March 2015 AERA consists of 153 autonomous radio-detection stations,
distributed with different spacings, ranging from 150~m in the dense core up to
750~m, covering an area of about 17~km$^2$.  Different types of antennas are
used, including logarithmic periodic dipoles and active bowtie antennas (named
"Butterfly antennas" \cite{Charrier:2012zz}) covering the frequency range from
30 to 80 MHz \cite{Abreu:2012pi, Abreu:2011fb}.

\section{Precision measurement of the radio emission in air showers}
\begin{figure}[t]\centering
\includegraphics[width=0.99\linewidth]{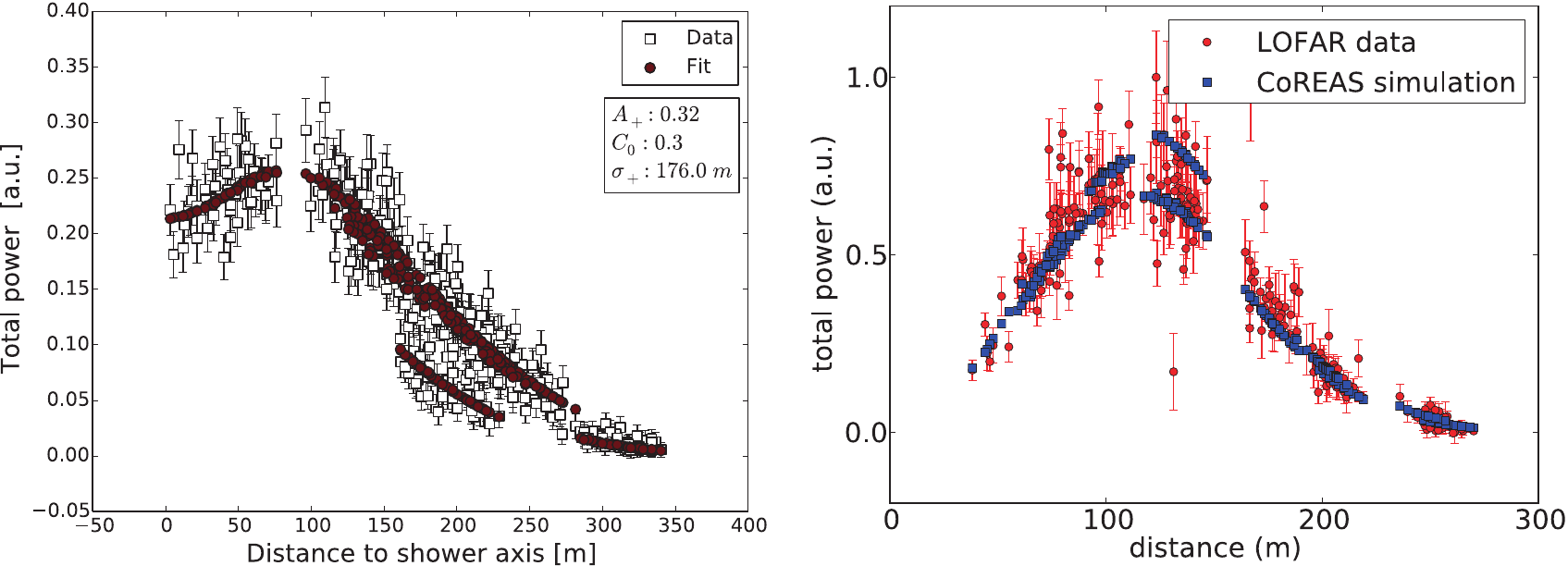}
\caption{Total power measured in an air shower as a function of the distance to
the shower axis (in the shower plane) as measured by LOFAR in the frequency
range $30-80$ MHz (\lleft) \cite{Nelles:2014gma} and $110-190$ MHz (\rright)
\cite{Nelles:2014dja}.}
\label{lat}
\end{figure}

LOFAR combines a high antenna density and a fast sampling of the measured
voltage traces in each antenna. This yields very detailed information for each
measured air shower and the properties of the radio emission have been measured
with high precision. At the Pierre Auger Observatory air showers are measured
simultaneously with various detector systems: radio detectors, fluorescence
light telescopes, water-\Cerenkov detectors, and underground muon detectors.
This unique combination yields complementary information about the showers and
allows to investigate correlations between the various shower components.
Some important aspects of radio emission in air showers are reviewed in the
following. We focus on radio emission in the frequency range $30-80$~MHz,
only one result (\fref{lat} \rright) deals with higher frequencies.

\begin{figure}[t]\centering
\includegraphics[width=0.99\linewidth]{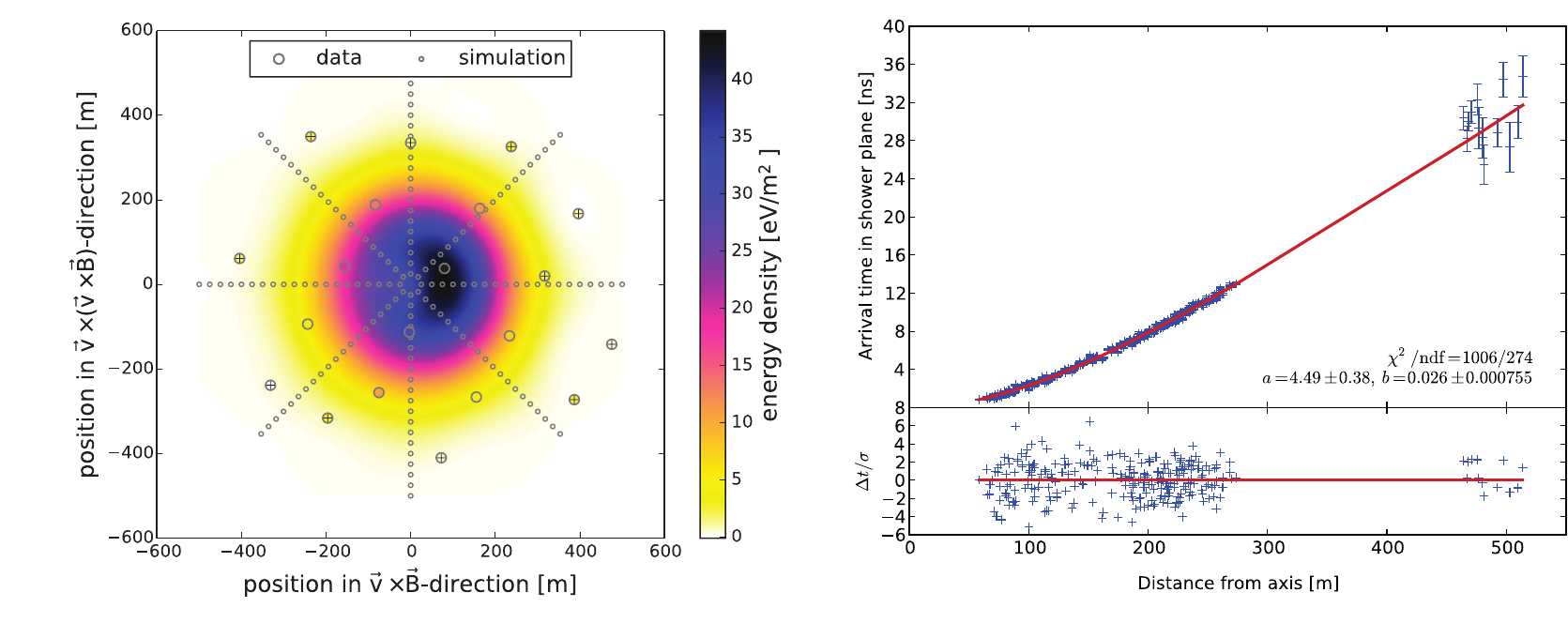}
\caption{\LLeft: Footprint of an air shower measured with AERA at the Pierre
Auger Observatory, the boxes representing the radio-detector stations. The
colored background represents predictions of the radio signal according to
simulations
\cite{SchulzIcrc15}. 
 \RRight: The arrival time of the signals, relative to a plane as a function to
the distance to the shower axis as measured with LOFAR.  The lower right graph
illustrates the arrival time differences with respect to a fit (hyperboloid)
\cite{Corstanje:2014waa}.  }
\label{footshape}
\end{figure}

\paragraph{Lateral distribution function of the radio signals}
The footprint of the radio emission recorded at ground level is not
rotationally symmetric \cite{Nelles:2014xaa,Nelles:2014gma,SchulzIcrc15}, such
as, e.g., the particle content of a shower, see \fref{footshape} (\lleft).
Radio emission is generated through interactions with the Earth magnetic field,
which yields a bean-shaped footprint on the ground.  The correct reference
system is in the shower plane (perpendicular to the shower axis), with one
coordinate axis along the $\vec v\times\vec B$ direction and the second axis
along the $\vec v\times (\vec v \times\vec B)$ direction.  Here $\vec v$ is
the propagation velocity vector of the shower (parallel to the shower axis) and
$\vec B$ represents the direction and strength of the Earth magnetic field. 
The measured power in the frequency range
$30-80$ MHz is plotted as a function of the distance to the shower axis in
\fref{lat} (\lleft). For example at a distance of 200 m from the shower axis,
ambiguities are visible in this one-dimensional projection: the recorded signal
strength is a function of the azimuth angle, which results in the visible
structure.  Historically, the lateral distribution of the radio signal on the
ground has often been parameterized with a simple exponential function (e.g.,
\cite{allanrev}).  However, the LOFAR measurements suggest that the radio
emission should be parameterized by a more complex expression: a
two-dimensional Gaussian function is used to describe the approximately
exponential fall-off at large distances from the shower axis. A second
(smaller) two-dimensional Gaussian function is subtracted from the first one to
describe the ring structure of the signal close to the shower axis. To
reproduce the observed bean shape, the centers of both Gaussian functions are
slightly offset.  
The power at position $(x',y')$ in the shower plane (perpendicular to the shower
axis) is described as
\begin{equation} \label{annaldf}
 P(x',y')=A_+ \exp\left(-\frac{(x'-X_+)^2+(y'-Y_+)^2}{\sigma_+^2}\right)
         -A_- \exp\left(-\frac{(x'-X_-)^2+(y'-Y_-)^2}{\sigma_-^2}\right) ,
\end{equation}
Where the parameters are the scaling factors $A_+$ and $A_-$ (where in general
$A_+>>A_-$), the width of the Gaussian functions $\sigma_+$ and $\sigma_-$, and
the centers of the two-dimensional Gaussian distributions $(X_+,Y_+)$ and
$(X_-,Y_-)$, respectively.

\paragraph{First quantitative measurements in the frequency range 120-240 MHz}
Radio emission from extensive air showers has also been recorded with the LOFAR
high-band antennas in the 200 MHz frequency domain \cite{Nelles:2014dja}. The
measured power is depicted in \fref{lat} (\rright) as a function of the
distance to the shower axis. A clear maximum is visible at distances around 120
m in this one-dimensional projection, indicating a clear (\Cerenkov) ring
structure. Such rings are predicted from theory: relativistic time compression
effects lead to a ring of amplified emission, which starts to dominate the
emission pattern for frequencies above $\sim100$ MHz. The LOFAR data clearly
confirm the importance to include the index of refraction of air as a function
of height into calculations of the radio emission.

\paragraph{Shape of the shower front}  
The precise shape of the radio wavefront is a long-standing issue. In the
literature different scenarios have been discussed: a spherical, conical, or
hyperbolic shape (e.g.\ \cite{Apel:2014usa}). The LOFAR findings clearly
indicate that a hyperboloid is the best way to describe the shape of the
measured wavefront \cite{Corstanje:2014waa}.  A hyperboloid asymptotically
reaches a conical shape at large distances from the shower axis and can be
approximated as a sphere close to the shower axis. A measured wavefront of a
shower registered with LOFAR is shown in \fref{footshape} (\rright).  The time
difference relative to a plane is plotted as a function of distance to the
shower axis.  The line indicates a fit of a hyperboloid to the measured data.
The lower part of the graph shows the time differences of the individual
antennas with respect to the fit function. 

\begin{figure}[t]\centering
\includegraphics[width=0.39\linewidth]{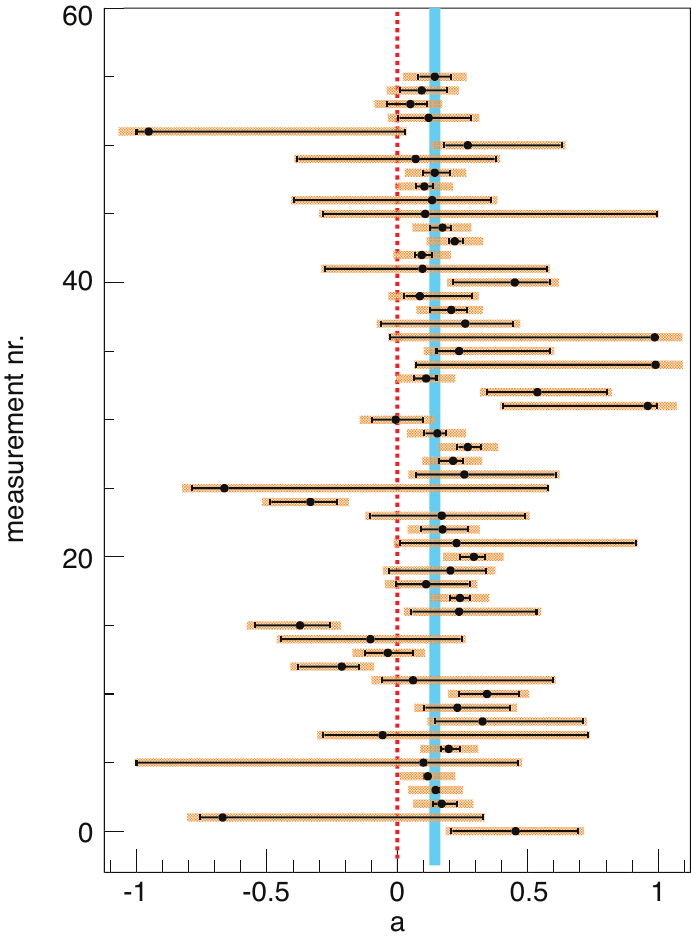}
\includegraphics[width=0.59\linewidth]{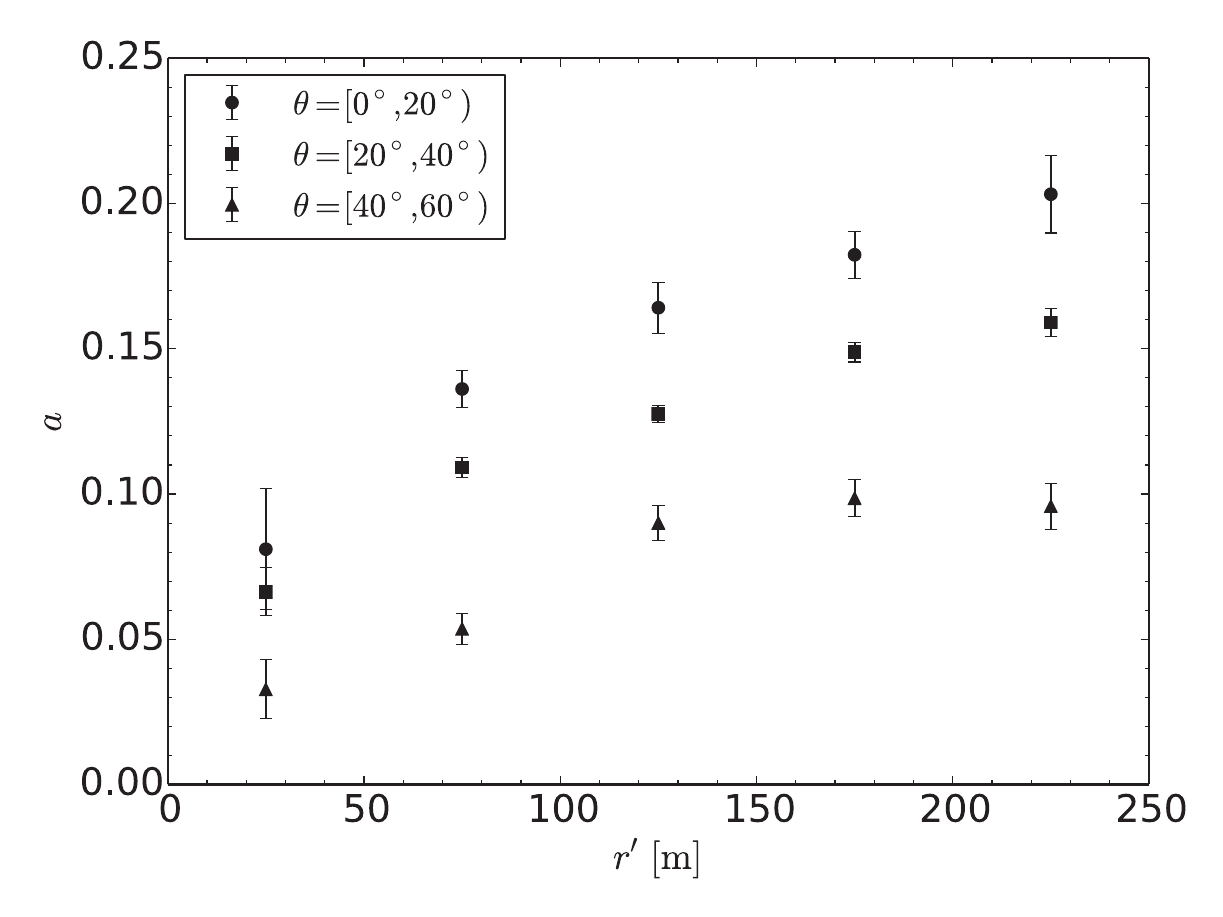}
\caption{Radio emission processes in the atmosphere. The ratio $a$ between
contributions due to the Askaryan effect and the geomagnetic emission is
plotted as measured at the Auger Observatory for different air showers (\lleft)
\cite{Aab:2014esa} and as a function of the distance to the shower axis (in the
shower plane) for showers with different zenith angles as measured by LOFAR
(\rright) \cite{Schellart:2014oaa}.}
\label{aparameter}
\end{figure}

\paragraph{Polarization of the radio signal}
The radio emission in extensive air showers originates from different
processes. The dominant mechanism is of geomagnetic origin
\cite{radionature,Ardouin:2009zp}. Electrons and positrons in the shower are
accelerated in opposite directions by the Lorentz force exerted by the magnetic
field of the Earth. The generated radio emission is linearly polarized in the
direction of the Lorentz force ($\vec v \times \vec B$), where $\vec v$ is the
propagation velocity vector of the shower (parallel to the shower axis) and
$\vec B$ represents the direction and strength of the Earth magnetic field.  A
secondary contribution to the radio emission results from the excess of
electrons at the front of the shower (Askaryan effect)
\cite{Marin:2011bga,Belletoile:2015rea}. This excess is built up from electrons
that are knocked out of atmospheric molecules by interactions with shower
particles and by a net depletion of positrons due to annihilation.  This charge
excess contribution is radially polarized, pointing towards the shower axis.
The resulting emission measured at the ground is the sum of both components.
Interference between these components may be constructive or destructive,
depending on the position of the observer/antenna relative to the shower. The
emission is strongly beamed in the forward direction due to the relativistic
velocities of the particles.  Additionally, the emission propagates through the
atmosphere, which has a non-unity index of refraction that changes with height.
This gives rise to relativistic time-compression effects, most prominently
resulting in a ring of amplified emission around the \Cerenkov  angle, see
\fref{lat}.  By precisely measuring the polarization direction of the electric
field at each antenna position, the ratio $a$ between the contributions from
the Askaryan effect and the geomagnetic emission is measured.  The resulting
ratios as measured at the Auger Observatory for individual air showers are
shown in \fref{aparameter} (\lleft) \cite{Aab:2014esa}.  An average value of
$a=14\%\pm2\%$ is obtained.  At LOFAR the ratio $a$ has been measured as a
function of the distance to the shower axis for showers with different zenith
angles, as depicted in \fref{aparameter} (\rright) \cite{Schellart:2014oaa}.
The figure illustrates that the contribution through the Askaryan effect
increases with increasing distance to the shower axis and it is more pronounced
for vertical showers (with small zenith angle).  

\paragraph{Confirmation of simulation codes} 
The detailed investigations of the properties of the radio emission and the
comparison between measurements and predictions from simulations demonstrate
that the simulation code CoREAS \cite{CoREAS} fully describes all relevant
features of the radio emission in air showers and the predictions can be used
to interpret the measured air-shower data.

\paragraph{Probing atmospheric electric fields during thunderstorms}
Radio detection of air showers can also be used for auxiliary science, such as
the measurements of electric fields in the atmosphere during thunderstorms
\cite{buitinkthunder,Apel:2013dga,Schellart:2015kga}.  The footprint of
the radio emission from an air shower, which developed during a thunderstorm is
shown in \fref{mass} (\rright) \cite{Schellart:2015kga}. The intensity and
polarization patterns of such air showers are radically different from those
measured during fair-weather conditions. The figure illustrates the
polarization as measured with individual LOFAR antennas (arrows) in the shower
plane.  An arrow labeled "normal" indicates the expected polarization direction
for fair-weather conditions.  LOFAR antennas are grouped into circular
stations, of which seven are depicted.  The position of the shower axis,
orthogonal to the shower plane, is indicated by the intersection of the dashed
lines. With the use of a simple two-layer model for the atmospheric electric
field, these patterns can be well reproduced by state-of-the-art simulation
codes. This in turn provides a novel way to study atmospheric electric fields.

\section{Measuring properties of cosmic rays with the radio technique}
The ultimate objective is to fully characterize the incoming cosmic ray with
the radio measurements, i.e., to determine its arrival direction, its energy,
and the particle type/mass (see, e.g., \cite{Apel:2014jol}).

\paragraph{Arrival direction}
A precise description of the shape of the shower front is essential to
correctly reconstruct the direction of the incoming cosmic ray.  Different
shapes have been investigated at LOFAR as discussed above
\cite{Corstanje:2014waa}. Reconstructing the arrival directions of the same
measured showers with different wavefront shapes results in differences in the
reconstructed arrival directions of typically less than 1\deg between a plane
and a hyperboloid and typically less than 0.1\deg between a cone and a
hyperboloid. Based on these investigations it is expected that the angular
resolution for the arrival direction of the shower is better than 1\deg.
Corresponding analyses at the Auger Observatory are under way, expecting a
similar precision.

\begin{figure}[t]\centering
\includegraphics[width=0.99\linewidth]{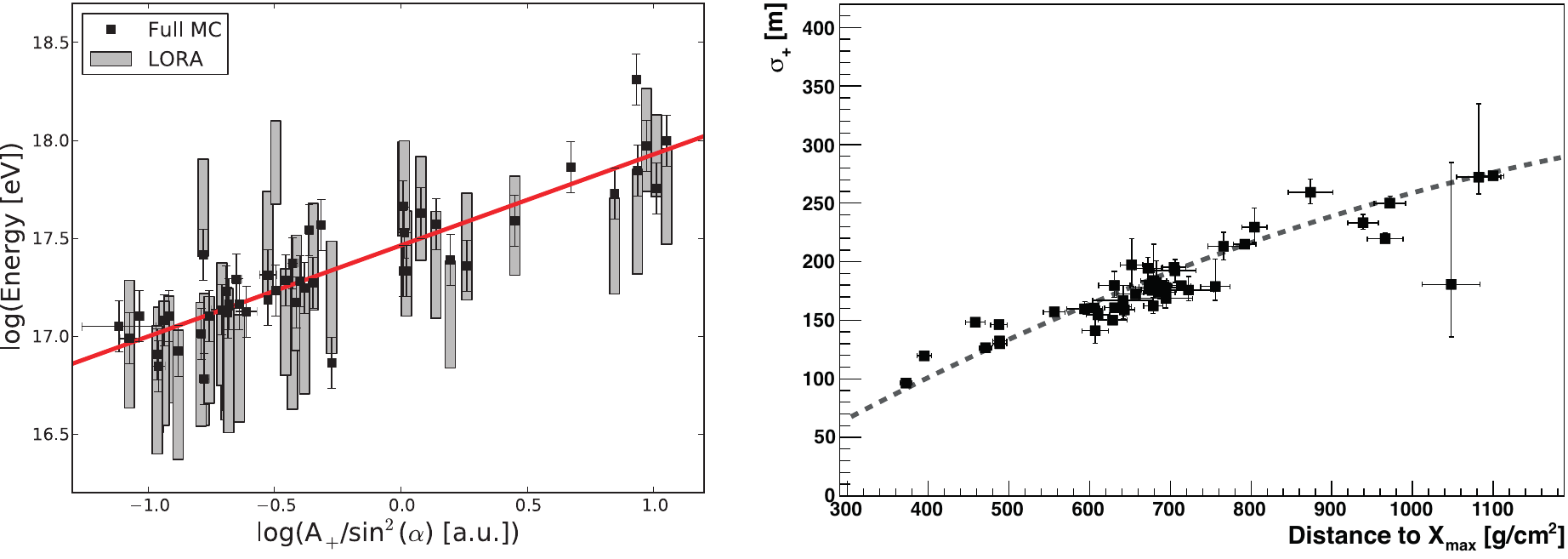}
\caption{Determining the properties of cosmic rays with radio measurements
\cite{Nelles:2014gma}.  \LLeft: Energy of the cosmic ray (determined from
simulations and measured with the LORA air shower array) as a function of the
integrated measured radio power.  \RRight: Width of the footprint of the radio
emission as a function of the distance to the shower maximum.}
\label{em}
\end{figure}

\paragraph{Energy}
The parameters of the function to model the intensity pattern of the radiation
on the ground (as described above) are sensitive to the properties of the
shower-inducing cosmic rays  \cite{Nelles:2014gma}.  The integral of the
measured power density is proportional to the shower energy. This is
illustrated in \fref{em} (\lleft). The shower energy is plotted as a function
of a parameter, that is proportional to the integrated power. The measured
signal strength is corrected for a factor, which depends on the angle between
the shower axis and the direction of the geomagnetic field. This is necessary,
since the dominant geomagnetic emission strongly depends on this angle. The
shower energy is determined in two ways in the figure: it is derived from Monte
Carlo simulations (using CORSIKA \cite{corsika} and CoREAS) and it has been
measured with the particle detector array at LOFAR (LORA). A clear correlation
can be seen between the shower energy and the measured radio intensity on the
ground.

A very sophisticated analysis is ongoing at the Auger Observatory
\cite{glaser-arena12,GlaserIcrc15,Aab:2015vta}.  The energy content of the
radio emission in air showers in the frequency range $30-80$~MHz has been
measured on an absolute scale.  A 1~EeV air shower, arriving perpendicularly to
the geomagnetic field radiates 15.8~MeV in the radio frequency regime.  Since
radio emission is not suffering from attenuation in the atmosphere and the
radiation strength can be calculated from first principles, measuring the radio
emission seems to be a very promising tool to establish the absolute energy
scale of air-shower detectors.

\paragraph{Depth of the shower maximum}
Several methods are studied at LOFAR and the Auger Observatory to derive the
type of the incoming cosmic ray or its atomic mass $A$ from the radio
measurements.  The key is to measure the depth of the shower maximum in the
atmosphere, which is proportional to $\ln A$.  The width of the measured
footprint is proportional to the geometrical distance from the antennas to the
position of the shower maximum. This correlation is shown in \fref{em}
(\rright).  This method is used at LOFAR and at the Auger Observatory to
estimate the cosmic-ray mass \cite{Nelles:2014gma,SchulzIcrc15}.

\begin{figure}[t]\centering
\includegraphics[width=0.99\linewidth]{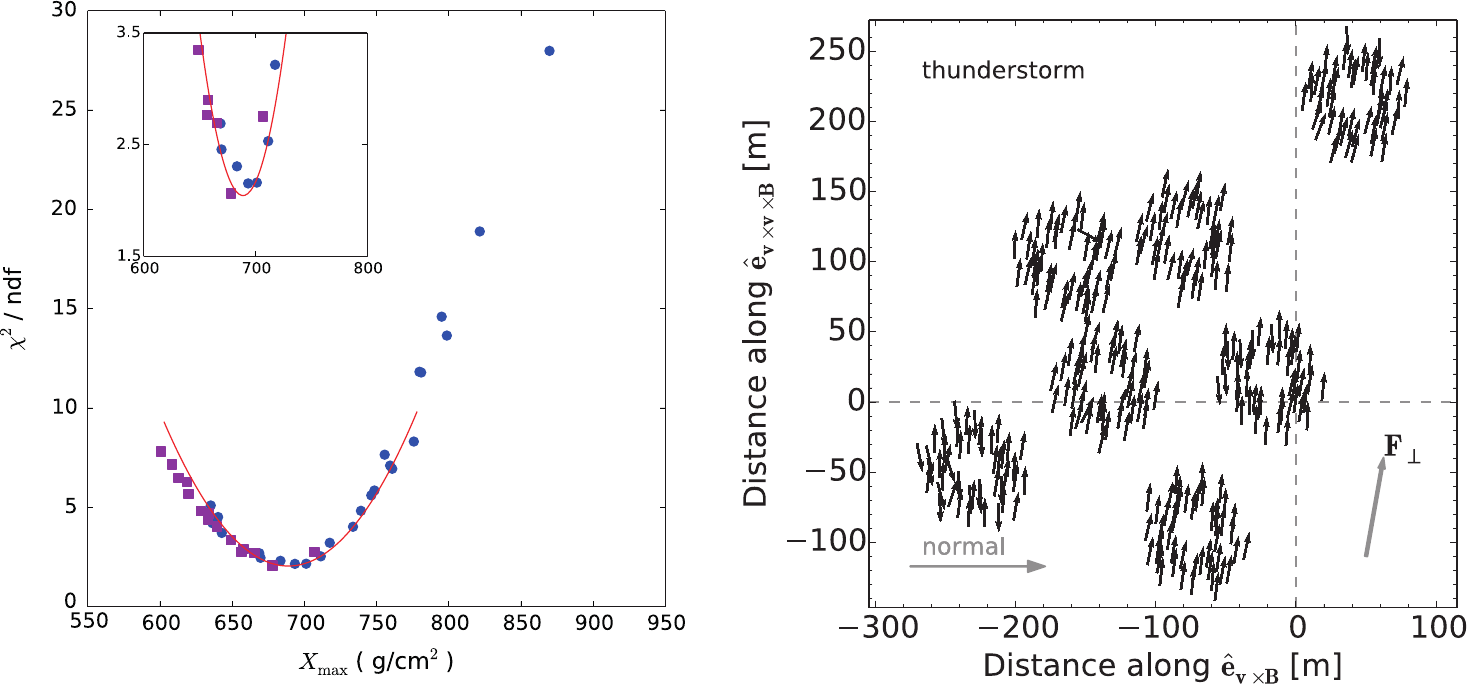}
\caption{\LLeft: Determining the depth of the shower maximum from the radio
measurements. The $\chi^2$ values of simulated showers are shown as a function
of the depth of the shower maximum \cite{Buitink:2014eqa}.
\RRight: Footprint of the radio emission of an air
shower, which developed during a thunderstorm \cite{Schellart:2015kga}.}
\label{mass}
\end{figure}

To obtain a precise value of the depth of the shower maximum for each shower,
we apply the following procedure, originally developed at LOFAR and now also
applied at the Auger Observatory \cite{Buitink:2014eqa,SchulzIcrc15}.  The
shower direction and energy are obtained from the measured signals in the
particle detectors and the radio antennas. With these parameters simulations
are initiated, simulating the development of air showers and the accompanying
radio emission, using the CORSIKA  and the CoREAS codes, for primary protons
and iron nuclei. Due to fluctuations in the individual particle interactions in
an air shower this results in cascades with a wide distribution for the depth of
the shower maximum (all of them having the same energy and direction of
incidence as the measured one). The predicted signals in the particle
detectors and the radio antennas are then compared to the measurements. A
$\chi^2$ method is used to determine the simulated shower which best matches
the measured values. This yields a value for the depth of the shower maximum.
The method is illustrated in \fref{mass} (\lleft). The $\chi^2$ values are
depicted as a function of the depth of the shower maximum \Xmax. A parabola is
fitted to the $\chi^2$ values and the minimum of this function gives the
estimated \Xmax value for the measured shower. With this method \Xmax is
determined with an accuracy of ~17~\gcm2 at LOFAR.  At the Auger Observatory
this method is used to compare the depth of the shower maximum as obtained from
the radio measurements with the \Xmax values measured with the fluorescence
detectors. Quantitative analyses of the achieved accuracy are under way.

At the Auger Observatory other mass-sensitive observables are also
investigated: the shape of the shower front is sensitive to the geometrical
distance to the shower maximum; the shape of the shower front is approximated
with a hyperboloid and the angle between the shower plane and the hyperboloid
(at large distances from the shower axis) is sensitive to \Xmax.  The spectral
shape of the measured radio signal is also sensitive to \Xmax
\cite{grebe-arena12}.  This could be a very promising method for a sparse radio
array, where the only measurable radio signals are detected in a small number
of radio detectors.

%\section*{References}
%\bibliographystyle{elsart-num-jrh}
%\bibliography{cr}

\end{document}